\documentstyle[12pt]{article}
\begin{document}
\setlength{\unitlength}{1mm}
{\hfill  JINR E2-94-246, June 1994 } \vspace*{2cm} \\
\begin{center}
{\Large\bf The Conical Singularity And Quantum Corrections To Entropy Of
 Black Hole}
\end{center}
\begin{center}
{\large\bf  Sergey N.~Solodukhin$^{\ast}$}
\end{center}
\begin{center}
{\bf Bogoliubov Laboratory of Theoretical Physics, Joint Institute for
Nuclear Research, Head Post Office, P.O.Box 79, Moscow, Russia}
\end{center}
\vspace*{2cm}
\begin{abstract}
For general finite temperature different from the Hawking one there appears
a well known conical singularity in the Euclidean classical solution of
gravitational
equations. The method of regularizing the cone by regular surface is used
to determine the curvature tensors for such a metrics. This allows one to
calculate the one-loop matter effective action and the corresponding
one-loop quantum corrections to the entropy in the framework of the
path integral approach of Gibbons and Hawking. The two-dimensional and
four-dimensional cases are considered. The entropy of the Rindler space
is shown to be divergent logarithmically in two dimensions and quadratically
in four dimensions that coincides with  results  obtained earlier.
For the eternal 2D black hole we observe
finite, dependent on the mass, correction to the entropy.
The entropy of the 4D Schwarzschild black hole is shown to possess an
additional
(in comparison with the 4D Rindler space) logarithmically divergent correction
which does not vanish in the limit of infinite mass of the black hole.
We argue that infinities of the entropy in four dimensions are renormalized by
the renormalization of the gravitational coupling.
\end{abstract}
\begin{center}
{\it PACS number(s): 04.60.+n, 12.25.+e, 97.60.Lf, 11.10.Gh}
\end{center}
\vskip 1cm
\noindent $^{ \ast}$ e-mail: solod@thsun1.jinr.dubna.su
\newpage
\baselineskip=.8cm

One of the interesting problems of the black hole physics is a
microscopic explanation as the states counting of the Bekenstein-Hawking
entropy of the
black hole which is in four dimensions proportional to the area of
the horizon. In quantum field theory one can define the "geometric
entropy"associated with a pure state and a geometrical region by considering
the
pure state density matrix, tracing over the  field variables inside the
region to form the density matrix which describes the state of the field
outside the region [1-3]. Taking
this region to be a sphere
in the flat space-time,
the recent numerical study [3] shows that the corresponding entropy
scales as the surface area of the sphere.
Thus, no gravity is present  but the entropy thus defined behaves typically for
the black hole.
In [3] it has also been  observed that this quantity is quadratically divergent
when the ultraviolet
cutoff (the size of the lattice), $\epsilon \rightarrow 0$. In two dimensions
[3,5-8],
the geometric entropy is logarithmically divergent $S={1 \over 6} \log {\Sigma
\over \epsilon}$, where $\Sigma$ is the size of a box to eliminate infrared
problems.

The somewhat related definition
of the {\it black hole} entropy by tracing states outside the horizon has
been suggested in [4].

It has been argued in [5] that quantum mechanical geometric entropy
is the first quantum correction to the thermodynamical entropy. In  flat space,
the appropriately defined geometric entropy of a free field is just the quantum
correction to the Bekenstein-Hawking entropy of Rindler space [5]. In the case
of the
black hole,
 the fields propagating in the region just outside the horizon give the
main contribution to the entropy [1]. For very massive black holes this region
is
approximated as flat Rindler space. Therefore, one can expect that for the
black hole
the corresponding quantum correction to the entropy is essentially the same as
for Rindler space.

In [8] it has been shown that all these results can be obtained by using the
known finite-temperature
expression for the renormalized $<T^0_0>$ in the Rindler space.

The main goal of this paper is to calculate the quantum correction to the
Rindler
and black hole entropy by means of the path integral approach of Gibbons and
Hawking [9]. We  consider at first the two-dimensional case and  extend
the results obtained to four dimensions.
 The one-loop effective action for matter
in D=2 and its divergent part in D=4 are well known and expressed
in terms of geometrical invariants constructed from curvature. For general
finite temperature different from the Hawking one there appears the well known
conical
singularity  in the classical solution of gravitational equations.
Therefore, we are faced with the problem of describing the geometrical
invariants for
manifolds with conical singularities. Fortunately, for the cases under
consideration
this can be performed. The extension of gravitational action to the
conical geometries  can also be found in [10].

\bigskip

Let us consider the canonical ensemble for the system of gravitational field
($g_{\mu\nu}$)
and matter ($\varphi$) under temperature $T=1/2\pi\beta$.
Then, the partition function of the system  is given as the Euclidean
functional integral
\begin{equation}
Z(\beta)=\int_{}^{}[{\cal D} \varphi] [{\cal D} g_{\mu\nu}] exp[-I(\varphi,
g_{\mu\nu})]
\end{equation}
where the integration is taken over all fields ($\varphi, g_{\mu\nu})$ which
are real
in the Euclidean sector and periodical with respect to imaginary time
coordinate
$\tau$ with period $2\pi \beta$. The action integral in (1) is a sum of pure
gravitational action and action of matter fields:
\begin{equation}
I(\varphi, g)=I_{gr}(g)+I_{mat}(\varphi, g)
\end{equation}
In the stationary phase approximation, neglecting contribution of thermal
gravitons,
one quantizes only matter fields considering metric as classical. Then,
in the one-loop approximation one obtains:
\begin{equation}
Z(\beta)=exp[-I_{gr}(g, \beta)-{ 1 \over 2} \ln det \Delta_{g}]
\end{equation}
where the leading contribution is given by the metric which is a classical
solution and periodical with period $2\pi \beta$
with respect to imaginary time $\tau$.
For arbitrary $\beta$ this metric is known to have conical singularity
\footnotemark\
\addtocounter{footnote}{0}\footnotetext{In the strict sense, this metric is no
more the solution
of the Einstein equations. Nevertheless, it  still
gives the main contribution to the functional integral (1) in the class of
metrics
with conical singularity.}. Therefore, in general,
the integration in (1) must include the metrics with conical singularities
as well [11]. As it is well known, there exists special Hawking inverse
temperature
$\beta_H$ for which the conical singularity disappears. It seems to be
reasonable
to calculate  thermodynamical quantities as energy and entropy for arbitrary
$\beta$ and then take the limit $\beta \rightarrow \beta_H$.

With respect to partition function (1) one can define the average of the
energy:
\begin{equation}
<E>=-{ 1 \over 2\pi} \partial_\beta \ln Z(\beta)
\end{equation}
and entropy
\begin{equation}
S=(-\beta \partial_\beta +1) \ln Z(\beta)
\end{equation}
Hence, for the Hawking temperature we have
\begin{equation}
S_{BH} =(-\beta \partial_\beta +1) \ln Z(\beta) |_{\beta=\beta_H}
\end{equation}
As it follows from (3), in the stationary phase approximation $Z(\beta)$ can be
represented by the effective action
\begin{eqnarray}
&&\ln Z(\beta)=-{\cal G}_{eff}(g), \nonumber \\
&& {\cal G}_{eff}(g)= I_{gr}(g) +{\cal G}_{eff} (\Delta ),
\end{eqnarray}
where ${\cal G}_{eff}(\Delta)= { 1 \over 2} \ln det \Delta_g$ is the one-loop
 contribution to the effective action
due to matter fields. Then, from (6) we obtain
\begin{equation}
S_{BH}=(\beta\partial_\beta-1){\cal G}_{eff} (g)|_{\beta=\beta_H}
\end{equation}
where ${\cal G}_{eff}(\Delta)$ is considered in the background classical metric
with
conical singularity. Inserting (7) into (8) we obtain that the total entropy
\begin{equation}
S_{BH}=S^{clas}_{BH}+S^q_{BH}
\end{equation}
is a sum of classical entropy:
\begin{equation}
S^{clas}_{BH}=(\beta\partial_\beta-1)I_{gr}
\end{equation}
and quantum corrections:
\begin{equation}
S^q_{BH}=(\beta\partial_\beta -1){\cal G}_{eff}(\Delta)
\end{equation}

Using this general formulae,  let us consider the case of two dimensional
gravity
interacting with scalar conformal matter:
\begin{equation}
I_{mat}=\int_{}^{}{ 1 \over 2} (\nabla \varphi)^2 \sqrt{g} d^2z
\end{equation}
Then, we get that
\begin{equation}
{\cal G}_{eff} (\Delta) ={1 \over 2} \ln det (- \Box)
\end{equation}
where $\Box=\nabla_\mu \nabla^\mu$ is the Laplace operator defined with respect
to
metric with conical singularity.

In two dimensions the description of      manifolds
 with conical singularity is essentially simplified.
We begin with the consideration of the simplest example of a surface
like that described by the  metric:
\begin{equation}
ds^2=d\rho^2+({1 \over \beta_H})^2\rho^2 d\tau^2
\end{equation}
This space can be considered as an Euclidean variant of 2D Rindler space-time.
Assuming that $\tau$ is periodical with period $2 \pi \beta$, let us consider
the
new variable $\phi=\beta^{-1} \tau$ which has period $2\pi$.
Then, (14) reads
\begin{equation}
ds^2=d\rho^2+\alpha^2\rho^2d\phi^2,
\end{equation}
where $\alpha=({\beta \over \beta_h})$. It is the  standard cone with
singularity
at $\rho=0$. When $\alpha=1 \ (\beta=\beta_H)$ the conical
singularity in (15) disappears.

\bigskip

Having calculated the scalar curvature $R_{con}$ for the conical metric (15)
let us approximate [12] the cone   by  a regular
surface determined in 3D Euclidean space by the equations: $x=\alpha\rho \cos
\phi$,
$y=\alpha \rho \sin \phi,$ $ z=\sqrt{1-\alpha^2} \sqrt{\rho^2+a^2}$, and with
the metric
\begin{equation}
ds^2={{\rho^2 +a^2\alpha^2} \over {\rho^2+a^2}} d\rho^2 +\alpha^2\rho^2 d\phi^2
\end{equation}
In the limit $a \rightarrow 0$, the metric (16) coincides with the metric of
the cone (15). Calculating curvature for (16) and taking off the regularization
($a \rightarrow 0$) for the scalar curvature of the cone (15) we obtain
(see also [13]):
\begin{equation}
R_{con}={2(\alpha-1) \over \alpha} \delta (\rho),
\end{equation}
where $\delta(\rho)$ is the delta-function defined with respect to the measure:

$$
\int_{0}^{+ \infty} \delta ( \rho) \rho d \rho=1
$$

This regularization of the cone allows us to use the  results
obtained for determinants of elliptic operators on regular surfaces
and then, taking approximation (16),
to obtain
the relevant expressions for the surfaces with conical singularities.

\bigskip

On general grounds, the one-loop effective action (13) contains the divergent
and finite parts [14]:
\begin{equation}
{\cal G}_{eff} (\Box) ={\cal G}_{inf}(\Box) +{\cal G}_{fin}(\Box),
\end{equation}
for which, neglecting boundary terms, we have
\footnotemark\
\addtocounter{footnote}{0}\footnotetext{Our convention for the curvature and
Ricci tensor is $R^\alpha_{\ \beta\mu\nu}=\partial_\mu \Gamma^\alpha_{\
\nu\beta}-...,$
and $R_{\mu\nu}=R^\alpha_{\ \nu\mu\alpha}$.}:
\begin{equation}
{\cal G}_{inf}(\Box)={ 1 \over 48\pi} \int_{}^{}R \sqrt{g} d^2z \log ({L \over
\mu})^2
\end{equation}
\begin{equation}
{\cal G}_{fin}(\Box)={ 1 \over 96\pi} \int_{}^{}R \Box^{-1} R \sqrt{g} d^2z
\end{equation}
where $L$ is an ultraviolet cut-off.
Let us proceed with the conical metric (15). As it follows from (17), the Euler
number for the cone is the following:
\begin{equation}
\chi={ 1 \over 4\pi} \int_{}^{}R_{con}\sqrt{g} d^2z=(\alpha-1)
\end{equation}
and, consequently, for the divergent part of the effective action we get
\begin{equation}
{\cal G}_{inf}(\Box)={ 1 \over 12} (\alpha-1) \log ({L \over \mu})^2
\end{equation}

\bigskip

In order to calculate the finite part (20), let us
consider the function $\psi_{con}$ which is the solution of the equation:
\begin{equation}
\Box \psi_{con}= R_{con}, \ \ R_{con} ={2(\alpha-1) \over \alpha} \delta (\rho)
\end{equation}
Eq.(23) can be rewritten as follows:
\begin{equation}
(\rho^{-1} \partial_\rho (\rho \partial_\rho) +{ 1 \over {\alpha^2 \rho^2}}
\partial^2_\phi)
\psi= {4\pi(\alpha-1) \over \alpha} \delta^2(x,x')
\end{equation}
where $\delta^2(x,x')$ is the two-dimensional $\delta$-function satisfying the
condition

$$
\int_{0}^{2\pi}d \phi \int_{0}^{+ \infty}\delta^2(x,x') \rho d\rho=1
$$

One can see that (24) is just the equation for the Green function. The solution
is well known:
\begin{equation}
\psi_{con}={2(\alpha-1) \over \alpha } \ln \rho + w,
\end{equation}
where $w$ is a harmonical function , $\Box w=0$. Taking into account (23), (25)
we obtain the following for the finite part (20):
\begin{eqnarray}
&&{\cal G}_{fin} (\Box)={ 1 \over 96 \pi} \int_{}^{}R_{con} \psi_{con} \sqrt{g}
d^2z
\nonumber \\
&&= {1 \over 12} {(\alpha-1)^2 \over \alpha} \ln \epsilon + { 1 \over
24}(\alpha-1)
w(0)
\end{eqnarray}
where $w(0)$ is the value  of the function $w$ at $\rho=0$.
One can see that the "finite" part of the effective action is really divergent
in the low limit of the integration over $\rho$. Therefore we introduced the
regularization: the distance $\epsilon$  to the top of the cone. In the limit
$\epsilon \rightarrow 0$  (26) is logarithmically divergent.
Thus, the complete one-loop  effective action on the cone takes the form
\begin{equation}
{\cal G}_{eff}(\Box)= {(\alpha-1) \over 12} \log ({L \over \mu})^2 +{ 1 \over
12}
{(\alpha-1)^2 \over \alpha} \log \epsilon + {(\alpha-1) \over 24}  w(0)
\end{equation}
We see that there are two types of divergencies on the cone. The first is
the ultraviolet divergency related to the cut-off of Feynman diagrams on the
energy $L$
(what is equivalent to the introduction of some minimal distance $L^{-1}$). The
other
divergency arises when the distance to the top of the cone goes to zero.
One can expect
that in the self-consistent renormalization procedure all the distances cannot
be smaller
than the fixed ultraviolet scale $L^{-1}$. From this point of view, the
identification
$\epsilon=L^{-1}$ is fairly natural. Moreover, this identification turns out to
be
necessary when we compare (27) with results obtained earlier for determinants
on the cone
by means of the $\zeta$-function [5] (see also [15]).
Indeed, assuming $\epsilon=L^{-1}$ in (27) we obtain
\begin{equation}
{\cal G}_{\infty}(\Box)= {1 \over 12} (\alpha- { 1 \over \alpha}) \log {L \over
\mu}
\end{equation}
that coincides with that of obtained in [5].

\bigskip

Now we can calculate the corresponding correction to the entropy.
Inserting (27) into (11) we observe that the second term in (27) doesn't
contribute to the entropy (for $\beta=\beta_H$) and the first term in (27)
leads
to
\begin{equation}
S^q_{BH}={ 1 \over 12} \log ({L \over \mu})^2= { 1\over 6} \log { \Sigma \over
\epsilon},
\end{equation}
where we assumed that $w(0)=0$. This coincides with result previously obtained
for the quantum correction to the entropy of 2D
Rindler space-time.

\bigskip

Some remarks concerning the ultraviolet divergency of entropy (29)
are in order. According to general
recipes of renormalization in the quantum field theory, one must add
the relevant counter-terms to the
"bare" action  in order to cancel the divergency.
In our case one must add the following term to the classical action:

$$
{ s \over 4\pi} \int_{}^{}R \sqrt{g}d^2z
$$
which doesn't affect the classical equations of motion but contributes to the
effective action: $\Delta {\cal G}_{eff}=s(\alpha-1)$ and entropy: $\Delta
S=s$. The total entropy occurs to be
finite but undefined. All this procedure seems to be reasonable.
The entropy is determined up to an arbitrary constant. The above cancellation
of the divergency
means only the renormalization of this additive constant which does not
influence
the physics and can not be determined from the experiment.

On the other hand, introducing the ultraviolet cut-off $L$ in
statistical description of
the  system we introduce a grain scale $L^{-1}$. This means that we define  an
elementary
state of the system characterized by the size $\epsilon=L^{-1}$. The situation
looks similar to the one that we have in the classical statistical physics
[16]. Quasiclassically,
one can define the number of states in the region of the phase space $(p,q)$
as $\Delta \Gamma ={\Delta p \Delta q \over (2\pi \hbar)^s}$ ( $s$ is number
of freedoms of the system) where $\epsilon=(2\pi\hbar)$ is a scale
characterizing
the elementary state of the system. Then entropy $S=\ln \Delta \Gamma$ is
divergent
when the Planck constant $\hbar \rightarrow 0$ ($\epsilon \rightarrow 0$).
Thus, the result (29)
can be interpreted as an indication that there must exist a fundamental
scale which plays the role of ultraviolet regulator and naturally characterizes
(like the Planck constant $\hbar$ in standard statistical physics)
the size of elementary state of quantum gravitational system in phase space
(concerning this, see also [17]).

\bigskip

Let us now consider the 2D black hole with metric written in the
Schwarzschild-like gauge:
\begin{equation}
ds^2=g(x)d\tau^2+{ 1 \over g(x)} dx^2
\end{equation}
The metric is supposed to be asymptotically flat: $g(x) \rightarrow 0$ if
$x \rightarrow +\infty$.
We assume that $\tau$ in (30) is periodical with period $2\pi\beta$. Consider
the new angle coordinate $\phi=\tau / \beta$ which has period $2\pi$. If one
introduces
a new radial
coordinate $\rho$:
\begin{equation}
\rho=\int_{}^{}{dx \over \sqrt{g(x)}},
\end{equation}
the metric (30) takes the form
\begin{equation}
ds^2=\beta^2 g(\rho) d\phi^2 +d\rho^2
\end{equation}

Let the metrical function $g(x)$ have zero of the first order at the point
$x=x_h$. In the Minkowski space this point is event horizon. Near the horizon
we have $g(x)=g'|_{x_{h}}(x-x_{h})$ for the metric function.  For $\rho$ (31)
we obtain
\begin{equation}
\rho={2 \over \sqrt{g'_{x_h}}} (x-x_h)^{ 1/2}
\end{equation}
in which it is assumed that the horizon is located at $\rho=0$.
For the function $g(\rho)$ in the vicinity of this point we get:
\begin{equation}
g(\rho) \approx {\rho^2 \over \beta^2_H}
\end{equation}
where $\beta_{H}= 2/g'_{x_h}$. The metric (32) can be rewritten in the form
\begin{equation}
ds^2=[({\beta \over \beta_H})^2 \rho^2 d\phi^2+ d\rho^2] +f(\rho)d\phi^2
\end{equation}
where we directly extract the cone part of the metric with singularity at
$\rho=0$.
The function $f(\rho)=\beta^2 g(\rho)-({\beta \over \beta_H})^2 \rho^2$ near
the
point $\rho=0$ behaves as $f(\rho) \sim \rho^4$. Now we may regularize the cone
part of the metric (35) as before (16), calculate the scalar curvature and then
take off the regularization $(a \rightarrow 0)$
\footnotemark\
\addtocounter{footnote}{0}\footnotetext{
At this last stage, the important point is
the behavior of the function $f(\rho) \sim \rho^4$. Due to this, the
cross-terms like $\delta(\rho) f(\rho)$
and $\delta(\rho)f'(\rho)$ do not contribute.}. At the end, we obtain the
following
result for the curvature:
\begin{equation}
R={2(\alpha-1) \over \alpha} \delta (\rho)+R_{reg}, \ \ \ \alpha={\beta \over
\beta_H}
\end{equation}
where the first term is the contribution due to the conical singularity, while
the second,
regular, term takes the form:
\begin{equation}
R_{reg}={g''_\rho \over g} -{ 1\over 2} ({g'_\rho \over g})^2,
\end{equation}
One can see that $R_{reg}$ (37) has at $\rho=0$ the finite value determined by
the term of the 4-th order in the expansion of $g(\rho)$ (or $f(\rho)$).

\bigskip

In order to find  one-loop quantum corrections in the background metric
(30), (32), we have to find the function $\psi$ satisfying the equation:
\begin{equation}
\Box \psi= {2(\alpha-1) \over \alpha} \delta(\rho)+R_{reg}
\end{equation}
where the Laplacian $\Box$ for the metric (32) reads
\begin{equation}
\Box=\partial^2_\rho +{g' \over 2g} \partial_\rho +{ 1 \over \beta^2 g}
\partial^2_\phi
\end{equation}
Assuming that $\psi$ is independent of $\phi$ we get that out of the point
$\rho=0$ the general solution of (38) is the following
\begin{equation}
\psi= \ln g + b \int_{\rho}^{\Lambda '}{d \rho \over \sqrt{g}},
\end{equation}
where $b$ and $\Lambda '>0$ are still arbitrary constants. In the limit
$\rho \rightarrow 0$ the Laplacian (39) coincides with the Laplacian for
the cone (24). Hence, in order to obtain the $\delta$-singularity
in the r.h.s. of eq.(38) the $\psi$ (40) has to  coincide with the
corresponding
solution for the cone, $\psi_{con}$, in the limit $\rho \rightarrow 0$:
\begin{equation}
\psi \rightarrow \psi_{con}={ 2(\alpha-1) \over \alpha} \ln \rho \ \ if
\ \ \rho \rightarrow 0
\end{equation}
Due to (34) we get for the leading terms of (40):
\begin{equation}
\psi=(2-b\beta_H) \ln \rho
\end{equation}
The condition (41) gives value of the constant $b=2 /\beta$. Finally, the
solution of (38) reads:
\begin{equation}
\psi= \ln g(\rho) +{ 2 \over \beta} \int_{\rho}^{\Lambda '}{d\rho \over
\sqrt{g(\rho)}}
\end{equation}
or, equivalently, in terms of the coordinate $x$:
\begin{equation}
\psi= \ln g +{ 2 \over \beta} \int_{x}^{\Lambda}{dx \over g(x)}
\end{equation}

\bigskip

It is worth observing that the renormalized energy density of the scalar field
in the space-time (30),
as it follows from (20), is:
\begin{equation}
<T^0_0>_{ren}={ 1 \over 48\pi}(2g''_x -g'_x\psi'_x +{ 1\over 2} (\psi'_x)^2)
\end{equation}
For $\psi (x)$ (40) this reads
\begin{equation}
<T^0_0>_{ren}= {1 \over 48 \pi}(2g''_x-{ 1 \over 2g} (g'^2_x-b^2))
\end{equation}
This expression can be obtained by integrating the conformal anomaly [18].
For $b=2/\beta$
this energy density
at the space infinity ($x \rightarrow \infty$)
\begin{equation}
<T^0_0>_{ren} \rightarrow { 1 \over 24\pi}({ 1 \over \beta})^2
\end{equation}
coincides with the energy density of massless bosons with temperature $T={ 1
\over 2\pi\beta}$.

It should be noted that the choice of  constant $b$ in (40) means
the choice of the quantum state of the scalar field in the space-time of
black hole. Therefore, the fact that this  constant is related to the
temperature
$\beta$ of the gravitational system seems to be natural: the thermal states
of the black hole
\footnotemark\
\addtocounter{footnote}{0}\footnotetext{This state is fixed when the
integration in (1)
is performed over the Euclidean manifolds with the cyclic Killing vector
$\partial_\tau$
with period $2\pi\beta$.}
and quantum field in the black hole space-time are the same.

On the other hand, we can see that (46) is divergent at the horizon $(x=x_h)$
for general $\beta$ and becomes regular only if $\beta=\beta_H$ (see also [8]).
Thus, the Hawking temperature
$\beta_H$ is distinguished also in the sense that only for this temperature
the renormalized energy density of the quantum field, being in the thermal
equilibrium
with the black hole, turns out to be finite at the horizon.
Really, the infinite energy density means that something singular can happen
at the horizon when the backreaction is taken into account. Therefore, for
$\beta \neq \beta_H$ the backreaction must be essential for  justifying the
semiclassical approximation
(when we consider (3) instead of
the functional integral (1)).

\bigskip

Before  calculating the quantum corrections to the entropy of the 2D black
hole,
one would like to have some concrete description of 2D gravity. The simplest
way is to use the string-inspired dilaton gravity with the following action
[19]:
\begin{equation}
I_{gr}=-\int_{}^{}d^2z \sqrt{g}[e^{-2\Phi} (-R+4(\nabla \Phi)^2+Q^2)+
4\nabla^\mu(e^{-2\Phi} \nabla_\mu \Phi)]
\end{equation}
where the last, boundary, term is added [20] in order to the on-shell action
(48) for the
flat space-time to satisfy the condition: $I_{gr}(g=1)|_{on-shell}=0$. Then,
from the field
equations we obtain
\begin{eqnarray}
&&ds^2=g(x)d\tau^2+{ 1 \over g(x)} dx^2, \ g(x)= 1-2m e^{-Qx} \nonumber \\
&&\Phi=-{Q \over 2}x
\end{eqnarray}
The action (48), considered on the solution (49), takes the form [20]
\begin{equation}
I_{gr}=\int_{}^{}d\tau dx[ \partial_x(e^{-2\Phi} \partial_{x}g)]
\end{equation}
Assuming that $\tau$ is periodical with period $2\pi\beta$, for (50) we obtain
\begin{equation}
I_{gr}=2\pi\beta [e^{-2\Phi} \partial_x g]_{+ \infty} -
4\pi {\beta \over \beta_H} [e^{-2\Phi}]_{x_h},
\end{equation}
where $x_h$ is the point of the horizon, $g(x_h)=0$, and $2/\beta_H=[\partial_x
g]_{x_h}=Q$.
However, this naive calculation of the action
doesn't take into account that for $\beta \neq \beta_H$ there exists a
conical singularity at $x=x_h$ with contribution (36) to the curvature.
This leads to additional term in the action:

$$
4\pi({\beta \over \beta_H}-1) [e^{-2\Phi}]_{x_h}
$$

Therefore, the action
(48) being considered on the classical metric (49) with $\beta \neq \beta_H$ is
as follows:
\begin{equation}
I_{gr}=2\pi\beta [e^{-2\Phi} \partial_x g]_{+ \infty} -
4\pi  [e^{-2\Phi}]_{x_h},
\end{equation}
Thus, the $\beta$-dependent terms, calculable on the horizon, are mutually
cancelled in (52)
and  for
the classical entropy of the black hole [20,21] we get:
\begin{equation}
S^{clas}=4\pi[e^{-2\Phi}]_{x_h}=8\pi m
\end{equation}
On the other hand, we obtain
$M=2mQ$ for the mass of the black hole.
Hence, the entropy (53) can be written as : $S^{clas}=2\pi \beta_H M$.

\bigskip

Now let us calculate  quantum corrections to (53) according to the above
considered
procedure.

{}From (36) we obtain that the Euler number for the metric (30), (32), when
$\beta \neq \beta_H$, is the sum
\begin{equation}
{ 1 \over 4\pi}\int_{}^{}R\sqrt{g} d^2z= \chi_{con}+\chi_{reg}
\end{equation}
where the first term on the r.h.s. of (54) is the contribution due to
the conical singularity while the second term is a regular contribution.
As before, we have that
\begin{equation}
\chi_{con}= ({\beta \over \beta_H}-1)
\end{equation}
To calculate the regular part, $\chi_{reg}$, it is convenient to use metric
in the form (30). Then,
 we obtain $R_{reg}=g''$ for the curvature. Consequently,
for the regular term in (54) one gets
\begin{eqnarray}
&&\chi_{reg}={ 1 \over 4\pi}\int_{}^{}g''dxd\tau \nonumber \\
&&=- {\beta \over \beta_H}
\end{eqnarray}
and the total Euler number (54)
turns out to be independent of $\beta$:
$\chi=-1$.
The divergent part of the one-loop effective action (19) is as follows:
\begin{equation}
{\cal G}_{inf}(\Box)=-{ 1 \over 12} \log ({L \over \mu})^2
\end{equation}
For the finite part of the effective action (20) we get
\begin{eqnarray}
&&{\cal G}_{fin} = { 1 \over 96\pi}
\int_{}^{}R \Box^{-1}R \sqrt{g}d^2z= {1 \over 96 \pi}\int_{}^{}R\psi
\sqrt{g}d^2z \nonumber \\
&&={\cal G}_{con}+{\cal G}_{reg}
\end{eqnarray}
where ${\cal G}_{con}$ is as follows:
\begin{equation}
{\cal G}_{con}={\beta \over 24} ({\alpha-1 \over \alpha})\int_{0}^{+\infty}
\delta ( \rho) \psi \sqrt{g(\rho)} d \rho
\end{equation}
The integrand in (59) is non-zero only at $\rho=0$ where $\psi=\psi_{con}$ and
$\sqrt{g}=\beta^{-1}_{H} \rho$. Hence, ${\cal G}_{con}$ coincides
with one we had for the cone (22):
\begin{equation}
{\cal G}_{con}={ 1 \over 12} {(\alpha-1)^2 \over \alpha} \ln \epsilon
\end{equation}
where the regularization, distance $\epsilon$ from the horizon ($\rho=0$), was
introduced.

The regular part  of (58)
\begin{equation}
{\cal G}_{reg}={\beta \over 48} \int_{x_h}^{+\infty}R_{reg} \psi dx
\end{equation}
does not contain divergencies in the low limit of the integration (the terms
like
$\epsilon \ln \epsilon$ vanish in the limit $\epsilon \rightarrow 0$).
In (61) we use the black hole metric in the form (30). For the concrete  metric
(49)
we get that

$$
\int_{x}^{\Lambda}{dx \over g(x)}=-{\beta_H \over 2} \ln g(x) -x +{ 1 \over Q}
\ln (e^{Q \Lambda}-2m)
$$

and $\psi (x)$ (44) takes the form
\begin{equation}
\psi=(1-{\beta_H \over \beta}) \ln g -{2x \over \beta} +{2 \over \beta Q}
\ln (e^{Q \Lambda}-2m)
\end{equation}
Inserting this into (61),  after the calculations we obtain
\begin{equation}
{\cal G}_{reg}={ \alpha \over 12} -{ 1 \over 12} \ln ({e^{Q \Lambda}-2m \over
2m})
\end{equation}

Collecting (60), (63) and (57), for the effective action we finally obtain:
\begin{equation}
{\cal G}_{eff}(\Box)=-{1 \over 12} \ln ({L \over \mu})^2 +
{ 1 \over 12} {(\alpha-1)^2 \over \alpha} \ln \epsilon + { \alpha \over 12}
-{ 1 \over 12} \ln ({e^{Q\Lambda}-2m \over 2m})
\end{equation}
where $\alpha={\beta \over \beta_H}$. Identifying $\epsilon=L^{-1}$,
for the infinite part of the effective action we obtain:
\begin{equation}
{\cal G}_{\infty}(\Box)=-{ 1 \over 12} {(\alpha^2+1) \over \alpha} \ln {L \over
\mu}
\end{equation}
As one can see, the infinite part does not depend on the concrete
form  of the black hole solution. Probably, the result (65) is worth checking
by means the
the alternative calculation, for example, with the  help of the
$\zeta$-function.
For the quantum correction to the entropy (53)  for $\beta=\beta_H$
we obtain:
\begin{equation}
S^q_{BH}={ 1 \over 6} \ln {\Sigma \over \epsilon} +
{ 1 \over 12} \ln ({e^{Q\Lambda}-2m \over 2m})
\end{equation}
The divergent part of (66) coincides with the quantum correction to the entropy
in the case of the Rindler
space-time (29). Obviously, this justifies the  approximation of the black hole
space-time near the horizon by the Rindler space, which was considered earlier
[1,5-8].

In terms of the classical mass $M$ and the Hawking temperature $\beta_H$ the
total
entropy (9) can be written as follows:
\begin{equation}
S_{BH}=2\pi\beta_H M +
{ 1 \over 12} \ln [{({2 \over \beta_H})e^{2\Sigma \over \beta_H}-M \over M}]
+{ 1 \over 6} \ln {\Sigma \over \epsilon}
\end{equation}
where we identified $\Lambda=\Sigma$.

In comparison with the Rindler space, for the black hole case we observe the
finite correction to entropy (67) which logarithmically depends on
the black hole mass $M$. This means, in particular, that the temperature
of the system defined as $T^{-1}=\partial_M S$ is no more $T_{H}$ but possesses
some corrections. This can be considered as an indication of that the
backreaction
must be taken into account. Indeed, the classical black hole solution
does not give the extremum of the semiclassical statistical sum (7). The
configuration, which is the minimum of the one-loop effective action ${\cal
G}_{eff}$,
must be considered. Generally, this quantum corrected configuration may
essentially
differ from the classical one [22-24]. In any case, such  thermodynamical
quantities
as temperature $\beta_H$, mass $M$ and entropy must be re-calculated.
Unfortunately, in general the quantum corrected field equations are not
exactly solvable. Recently [25-27], this has been considered for the RST model
where
the exact solution is known. It should be noted, however, that the backreaction
may change only the finite part of the entropy while its divergent part, as it
follows
from our consideration, remains unchanged.

\bigskip

Let us now  apply our method to the 4D case.
Assume that the gravitational field in four dimensions is described by
the standard Einstein-Hilbert action:
\begin{equation}
I_{gr}={ 1 \over 16\pi\kappa} \int_{}^{}d^4x \sqrt{g} R^{(4)} + \ \ boundary \
\
terms
\end{equation}
where the gravitational constant $\kappa$ has dimensionality of length squared
$[l^2]$.

The Rindler space in four dimensions is described by the metric
\begin{equation}
ds^2={\beta^2 \over \beta^2_H} d\phi^2 +d\rho^2 +dx^2+dy^2
\end{equation}
which for $\beta \neq \beta_H$ can be represented as a direct product of
the two-dimensional cone (15) on the 2D plane: $C^2 \otimes R^2$. Applying the
regularization
procedure (16) to the cone part of the metric (69), we obtain that the 4D
scalar curvature
for  (69) in the limit $a \rightarrow 0$ coincides with the curvature of the
2D cone (17):
\begin{equation}
R^{(4)}={2(\alpha -1) \over \alpha}\delta (\rho), \ \ \alpha={\beta \over
\beta_H}
\end{equation}

We are also interested in the spherically symmetric metric describing the 4D
black hole
\begin{equation}
ds^2=\beta^2 g(\rho) d\phi^2+d\rho^2+r^2(\rho)(d\theta^2+\sin^2 \theta
d\varphi^2)
\end{equation}
Near the horizon we have: $g(\rho)={\rho^2 \over \beta^2_H}$ and $r(\rho)=r_h+
{\rho^2 \over \beta_H}$, where $r_h$ is the value of the radius $r$ at the
horizon.
For $\beta \neq \beta_H$ there again exists the conical singularity at the
horizon ($\rho=0$).
The part of the metric (71) in the plane ($\phi, \rho$) coincides with the 2D
metric (32), (35).
Regularizing the conical singularity at $\rho=0$ as before
we obtain that the complete Riemann tensor is a sum of the regular
part (which is non-singular in the limit $a \rightarrow 0$) and the part coming
from the cone:
\begin{equation}
R^\mu_{\ \nu\alpha\beta}=
R^\mu_{con \ \nu\alpha\beta}+
R^\mu_{reg \ \nu\alpha\beta}
\end{equation}
The only non-trivial component of the contribution from the cone in (72) is
the following (for finite $a$):
\begin{equation}
R^\phi_{con \ \rho \phi \rho}={a^2(1-\alpha^2) \over
(\rho^2+a^2)(\rho^2+a^2\alpha^2)}
\end{equation}
Though the whole consideration can be generalized, we study here only the case
when
the regular part of the metric is Ricci-flat ($R^{reg}_{\mu\nu}=0$), i.e. it is
the solution of the Einstein equation in vacuum. Then, we obtain from (72),
(73) that
the scalar curvature for the metric (71) in the limit $a\rightarrow 0$ is also
given by
expression (70).

\bigskip

The divergent part of the one-loop effective action for scalar matter
described by the action

$$
I_{mat}={ 1 \over 2} \int_{}^{}(\nabla \varphi)^2 \sqrt{g}d^4x
$$

in four dimensions
(neglecting the boundary terms) takes the form (see, for example, [28]):
\begin{eqnarray}
&&{\cal G}_{inf}= { 1 \over 2} (\log det \Box)_{\infty} \nonumber \\
&&=-{ 1 \over 32 \pi^2}(B_0 L^4 +B_2 L^2 +B_4 \log ({L \over \mu})^2 )
\end{eqnarray}
where $L$ is the ultraviolet cut-off. The coefficients $B_k$ in (74) take the
form
(we omit the overall irrelevant coefficients dependent on the type of matter):
\begin{eqnarray}
&&B_0={1 \over 2}\int_{}^{}\sqrt{g} d^4x \nonumber \\
&&B_2=-{ 1 \over 6}\int_{}^{}R^{(4)} \sqrt{g}d^4x \nonumber \\
&&B_4=\int_{}^{}({1 \over 72} R^2 -{1 \over 180} R_{\mu\nu}R^{\mu\nu}
+{ 1\over 180}R^{\mu\nu}_{\ \ \alpha\beta} R_{\mu\nu}^{\ \ \alpha\beta}-
{ 1 \over  30} \Box R) \sqrt{g} d^4x
\end{eqnarray}

Considering   (74) on the Rindler background (69), and using (70) we obtain
\begin{equation}
{\cal G}_{inf}={1 \over 48 \pi} (\alpha-1)A_h L^2 -{\alpha  \over 64 \pi^2}V
L^4+
(\alpha-1)^2 A_h T(a,\alpha) \log {L  \over \mu}
\end{equation}
where $A_h=\int_{}^{}dxdy$ is the area of the Rindler horizon; $V$ is the
volume
of the space
(69) if $\beta=\beta_H$.
As one can see, $B_4$ is quadratic in curvature. For finite $a$ it gives the
last term
in the effective action (76) with the function $T(a,\alpha)$ having the form
\begin{equation}
T(a,\alpha)={1 \over a^2 } T(\alpha)
\end{equation}
where $T(\alpha)$ is a nonsingular function which takes finite value at
$\alpha=1$.
Thus, we obtain an additional (to the ultraviolet) divergency when we take
limit
$a\rightarrow 0$. Fortunately, the last term in (76) is proportional
to $(\alpha-1)^2$ and does not contribute to the energy (4) or entropy (6)
calculable at $\beta=\beta_H$.

{}From (76) we  get  the quantum correction to the entropy:
\begin{equation}
S^q_{BH}={1 \over 48\pi} {A_h \over \epsilon^2}
\end{equation}
where  the ultraviolet distance $\epsilon =L^{-1}$ was introduced.

The result (78) is exactly the one  obtained
for the geometrical entropy [1,3].
One can give this the following interpretation. Though we start from the flat
space-time,
considering system at the finite temperature $\beta$, we obtain the statistical
system in the effective 4D Euclidean space with the conical singularity.
Therefore,
the induced gravitational effects of the curvature play the role leading to the
non-trivial effective action (74) and entropy (78).

Notice that only the  contribution coming from the scalar curvature $R^{(4)}$
in the first power in effective action (74) leads to the quantum correction to
the
entropy of the Rindler space. This term takes the same form as the classical
("bare") gravitational action (68)  but with the $L^2$-divergent coefficient.
The two-dimensional example learns us that an extra ultraviolet divergencies
(in the limit $\epsilon \rightarrow 0$) can come from the "finite"
non-local terms in the complete effective action which are omitted in (74).
These terms are not exactly known in four dimensions. By means of methods
different from
ours, there recently appeared
results on the heat kernel asymptotic expansion on the curved cone [29].
They allow one  to obtain {\it all} divergencies  due to the conical
singularity:
which could come both  from the infinite and finite parts of the complete
effective action.
Comparing our result for the Rindler space (76) with that of [29] we observe
that
divergency coming from the finite part is proportional to $(\alpha-1)^2 L^2$.
Renormalizing the infinities of (74), (76) we introduce the same counter-terms
as for the manifolds without conical singularities. Thus, to renormalize the
$L^2$-divergency of (74), (76) it is enough to renormalize the gravitational
constant [30]:
\begin{equation}
\kappa^{-1}=\kappa^{-1}_B+L^2
\end{equation}
On the other hand, we must add new (absent in the regular case) local
counter-terms (cf. [31]), determined on the horizon surface, in order to
absorb the additional divergencies coming from the finite terms  in the
effective
action. However, this divergency is proportional to $(\alpha-1)^2$ and hence
does not
contribute to the entropy. Therefore, the renormalization of the gravitational
constant (79) is enough to renormalize the ultraviolet divergency of the
quantum correction to the entropy (78).

In the recent interesting preprint Susskind and Uglum [32] have also calculated
the
quantum correction to the entropy of the Rindler space which they consider as a
infinite mass limit of the black hole space-time.
In particular, it has also been observed that quantum correction to the entropy
is equivalent to
 the quantum correction to the gravitational constant
(for the discussion of this point, see also in [33]).

The metric (69) for $\alpha \neq 1$ is similar to the metric of a cosmic
string.
In the cosmic string interpretation of the metric (69) our procedure of
regularizing the conical singularity has the natural physical
justification. It means that we consider the string with the finite radius
"$a$" of
the kernel. This description is more realistic while the infinitely thin
cosmic string  (in the limit $a \rightarrow 0$) is an idealization.
Therefore, we could consider the parameter "$a$" in our above consideration as
a "phenomenological" one which is small but finite. This assumption allows us
to avoid an additional divergency in the effective action related with
the limit $a \rightarrow 0$.

\bigskip

Consider now the black hole described by the Schwarzschild solution.
The metric takes the form (71). Near the horizon ($\rho=0$) we have:
\begin{equation}
g(\rho)={\rho^2 \over \beta^2_H}-{4\rho^4 \over 3\beta^4_H},
\end{equation}
where $\beta_H=4M$, $M$ is the mass of the black hole.

The classical Bekenstein-Hawking entropy of the black hole is
well known:
\begin{equation}
S^{clas}_{BH}={A_h \over 4\kappa}
\end{equation}
where $A_h=4\pi r^2_h$ is the area of the horizon sphere; for the
Schwarzschild solution one has: $r_h=2M$.

Calculating the quantum correction to this entropy we observe the new point in
comparison with the Rindler case. Though the regular part of the metric is
Ricci-flat, the Riemann tensor $R^{\mu\nu}_{reg \ \alpha\beta}$ is non-zero.
{}From (72) we obtain that the term
\begin{equation}
R^{\mu\nu}_{reg \ \alpha\beta}R_{con \ \mu\nu}^{\alpha\beta}
\end{equation}
contributes non-trivially to  $B_4$ and to the effective action.
The conical Riemann tensor in (82) is proportional to $(\alpha-1)$ and
hence (82) leads to an additional correction to the entropy of the black hole.

In the limit $a \rightarrow 0$, the conical Riemann tensor $R^{\mu\nu}_{con \
\alpha\beta}$
is proportional to the $\delta$-function $\delta(\rho)$. Hence,  only the value
of
the regular Riemann tensor at the horizon is essential when we integrate
(82). From (71) and (80) we obtain:
\begin{equation}
R^\phi_{reg \ \rho \phi \rho} (\rho=0)={4 \over \beta^2_H}
\end{equation}
Substituting (72), (73) and (83) into the expression for $B_4$ (75)
in the limit $a \rightarrow 0$ we have:
\begin{equation}
B_{4}={32 \pi \over 270} {A_h \over \beta^2_H}{(1-\alpha)(\alpha^2+\alpha+1)
\over \alpha^2} +{(\alpha-1)^2 \over a^2}
T(\alpha)      +\alpha B_4^0
\end{equation}
where $B_4^0$ is the coefficient $B_4$ (75) calculated for the Schwarzschild
solution
if $\beta=\beta_H$.

The infinite part of the one-loop effective action (74) then takes the form:
\begin{eqnarray}
&&{\cal G}_{inf}={(\alpha-1) \over 48\pi}A_h L^2 -{\alpha V L^4 \over 64 \pi^2}
+{(\alpha-1)(\alpha^2+\alpha+1) \over 2160 \pi \alpha^2} {A_h \over M^2} \log
{L \over \mu}
\nonumber \\
&&+{(\alpha-1)^2 \over a^2} T(\alpha) \log {L \over \mu} -{\alpha \over
16\pi^2}
B_4^0 \log {L \over \mu}
\end{eqnarray}
Finally, for the quantum correction to the entropy we get
\begin{equation}
S^q_{BH}={A_h \over 4} ({1 \over 12\pi \epsilon^2} + {1 \over 180 \pi M^2}
\log {\Sigma \over \epsilon}),
\end{equation}
where the ultraviolet distance $\epsilon=L^{-1}$ was introduced.
The entropy (86) is proportional to the horizon area as before.  However, in
comparison with the Rindler case we observe additional, logarithmically
divergent, term in (86) which is dependent on the mass of the black hole.

Considering the entropy per the horizon area in the limit of the infinite black
hole
mass ($M\rightarrow \infty$), we obtain the entropy for the Rindler space.
This probably could justify the approximation of the infinite mass black hole
by
the Rindler space [1,32]. However, since the horizon area for the
Schwarzschild solution is $A_h=16\pi^2 M^2$, we observe that the logaritmically
divergent term in the complete entropy
\begin{equation}
S^q_{BH}={A_h \over 48 \pi \epsilon^2} +{1 \over 45} \log {\Sigma \over
\epsilon}
\end{equation}
is independent of the mass. It takes the form which is very similar to that
we had in the two-dimensional case (see (67)). The reason of different results
for the Rindler space and the black hole lies obviously in the different
topology of these manifolds. The topological numbers (like the Euler one)
vanish for the flat Rindler space while they are non-zero
for the  black hole and independent of the black hole mass.

To renormalize the $L^2$ and $\log L$ divergencies, in (85) we must add to the
bare gravitational action not only the Einstein-like term but also the
term $\kappa_1 B_4$
quadratic in
curvature
with new coupling constant $\kappa_1$.
The comparison with the exact results [29] shows that divergencies (both
$L^2$ and $\log L$), additional to (85) and coming from the "finite" terms in
the
complete effective action, are again proportional to $(\alpha-1)^2$ and they do
not
contribute to the entropy. Thus, we again obtain that the infinities of entropy
(86) (but not of effective action!) are renormalized by the renormalization
of only the coupling constants $\kappa$ and $\kappa_1$.

\bigskip

Finally, several remarks are in order of discussion. As it has been noted in
[32], only
the quantum corrections but not the classical entropy have
a clear interpretation in terms of the counting the states. To overcome this,
we may start from a zero bare gravitational action, assuming that the whole
gravitational dynamics is determined by an induced matter effective action.
Then, roughly speaking, the whole entropy of the black hole is a quantum
correction. An interesting example of the induced gravity is given by
the superstring theory (see also [34]) which is probably free from ultraviolet
divergencies. In the string theory, the space-time metric is not a primary
object. It appears in the low-energy approximation as a "quantum condensate"
of string excitations at energies $E << (\alpha')^{-{1 \over 2}}$ (see, for
example
[35]). Therefore, considering the low-energy effective action of the
string, we obtain that already the "classical" entropy can be identified with
the logarithm of appropriately counted number of such  string states.
However, this speculation needs further detailed investigation.

\bigskip

I am very grateful to Dima Fursaev for numerous remarks and criticism.
I also would like to thank Leo Avdeev, Evgenue Donets and Misha Kalmykov
for very useful
discussions. This work was supported in part by the grant RFL000 of
the International Science Foundation.

\end{document}